\documentclass[11pt,a4paper]{article}
\usepackage[margin=1in]{geometry}
\usepackage{amsmath,amssymb}
\usepackage{graphicx}
\usepackage{booktabs}
\usepackage{cite}
\usepackage{hyperref}

\usepackage[T1]{fontenc}
\usepackage{graphicx}
\usepackage{amsmath}
\usepackage{amssymb}
\usepackage{booktabs}
\usepackage{multirow}
\usepackage{url}    
\usepackage{cite}
\usepackage{array}
\usepackage{tabularx}
\usepackage{float}
\usepackage{titlesec}
\titlespacing*{\section}{0pt}{3.33ex}{2.19ex}
\titlespacing*{\subsection}{0pt}{3.09ex}{1.43ex}
\begin{document}

\title{DINO-A: Adapting Self-Distillation Vision Transformers to General Audio Representation Learning}

\author{
Tomasz~Radzikowski \and Mateusz~Modrzejewski \and Przemysław~Rokita \\[0.5em]
\normalsize Warsaw University of Technology, Warsaw, Poland \\
\normalsize \texttt{\{tomasz.radzikowski, mateusz.modrzejewski, przemyslaw.rokita\}@pw.edu.pl}
}

\date{}

\maketitle

\begin{center}
\textit{Preprint. Accepted at ICAISC 2026. Final version to appear in Springer Lecture Notes in Artificial Intelligence.}
\end{center}
\begin{abstract}
We present DINO-A, an adaptation of self-distillation from vision to general audio representation learning. While DINO has become a canonical method in self-supervised vision and prior audio work has explored latent prediction (BYOL-A) and masked modeling (Audio-MAE, BEATs), no prior work has brought canonical DINO to general audio classification in the way BYOL-A brought BYOL. DINO-A retains DINO's multi-crop, EMA teacher, and high-dimensional projection, replacing only the input modality and augmentations with log-mel spectrograms and the BYOL-A v2 augmentation block. We pretrain three backbones, two Vision Transformers with $8 \times 8$ and $16 \times 16$ patches and a convolutional encoder, on FSD50K and evaluate them with linear probing on ESC-50, Speech Commands v2, UrbanSound8K, and GTZAN. Three findings characterize the resulting representations. Patch resolution within the Vision Transformer family has consistent effect on representation quality, with smaller patches winning across all four tasks. The choice between Vision Transformer and convolutional backbone interacts with task type: convolutional networks lead on speech while Vision Transformers lead on environmental sounds and music. Under identical pretraining and evaluation conditions, DINO-A and BYOL-A v2 differ by 11.96 percentage points on average, and we trace this difference to two mechanisms: the interaction between DINO's high-dimensional projection space and FSD50K's limited scale, and the additional cost of multi-crop augmentation, which DINO uses but BYOL-A v2 does not. The high-dimensional projection space, central to DINO's success in vision, becomes a liability at FSD50K scale.
\end{abstract}

\section{Introduction}
\label{sec:intro}

Self-supervised learning (SSL) is now widely adopted for learning audio representations from unlabeled data, providing pre-trained encoders competitive with supervised baselines across environmental sound classification, music genre recognition, and speech-related tasks. Three main directions characterize current audio SSL: contrastive learning, latent prediction, and masked modeling, with representative methods including BYOL-A~\cite{niizumi2021byola,niizumi2023byola2} for latent prediction, Audio-MAE~\cite{huang2022audiomae} for masked reconstruction, and BEATs~\cite{chen2023beats} for masked discrete-token prediction.

Self-distillation, a fourth paradigm, has been comparatively underexplored for general audio. In computer vision, DINO~\cite{caron2021dino} demonstrated that a student-teacher framework with cross-entropy over a high-dimensional projection space, combined with multi-crop augmentation, produces representations that rival contrastive methods and exhibit emergent properties such as object segmentation through attention maps. The natural question is whether this paradigm transfers to general audio under the well-established adaptation pattern that produced BYOL-A from BYOL~\cite{grill2020byol}. Two prior works have applied self-distillation in audio sub-domains. DinoSR~\cite{liu2023dinosr} combines self-distillation with masked language modeling and online clustering for speech representation learning, while ASiT~\cite{atito2024asit} extends DINO with group-masked modeling for audio event classification. Neither performs a direct adaptation of canonical DINO to general audio under the BYOL-A pattern, nor a direct comparison between self-distillation and latent prediction on identical pretraining data.

We adapt DINO to general audio classification, following the BYOL-to-BYOL-A pattern. DINO-A keeps the canonical components of DINO (multi-crop, EMA teacher, centering and sharpening, high-dimensional projection) and replaces only the input modality and the augmentation pipeline, which we adopt from BYOL-A v2. We pretrain three backbone variants on FSD50K~\cite{fonseca2022fsd50k}, two Vision Transformers with patch sizes 8$\times$8 and 16$\times$16 and a convolutional encoder, and evaluate them via linear probing on four standard benchmarks. Matching the BYOL-A v2 pretraining data and evaluation protocol allows us to compare self-distillation and latent prediction directly, with the choice of self-supervised algorithm as the only variable.

Patch resolution within the Vision Transformer family has consistent impact on representation quality, with smaller patches winning across all four downstream tasks. The choice between Vision Transformer and convolutional backbone interacts with task type: CNNs win on speech, ViTs win on environmental sounds and music. DINO-A trails BYOL-A v2 by 11.96 percentage points on average under matched conditions, and we attribute this gap primarily to the interaction between DINO's high-dimensional projection space and the limited scale of FSD50K, with multi-crop augmentation introducing an additional structural cost that DINO-A bears but BYOL-A v2 does not. This points to AudioSet-scale pretraining as a natural extension of this work.

\section{Related Work}
\label{sec:related}

\paragraph{Self-supervised learning for general audio.}
Audio SSL has matured along three main lines, each producing competitive pretrained encoders for downstream classification. BYOL-A~\cite{niizumi2021byola,niizumi2023byola2} adapts the latent-prediction framework of BYOL~\cite{grill2020byol} to log-mel spectrograms, using cosine similarity in a low-dimensional projection space and a custom audio augmentation block (Mixup-BYOLA, RandomResizeCrop, RunningNorm). Audio-MAE~\cite{huang2022audiomae} extends masked autoencoders to spectrograms with mean-square reconstruction loss over masked patches. BEATs~\cite{chen2023beats} replaces reconstruction with discrete label prediction over an iteratively-refined acoustic tokenizer. Methods that depend on rich, fine-grained learning signals such as Audio-MAE and BEATs are typically pretrained on AudioSet~\cite{gemmeke2017audioset}, whereas BYOL-A achieves competitive performance on the smaller FSD50K~\cite{fonseca2022fsd50k}. We compare DINO-A directly with BYOL-A v2 under matched FSD50K pretraining, since the latter is the most established baseline for the BYOL-to-BYOL-A adaptation paradigm we follow.

\paragraph{Self-distillation in computer vision.}
DINO~\cite{caron2021dino} introduced a self-distillation framework in which a student network learns to match the output distribution of an exponentially-moving-average teacher across augmented views, using cross-entropy over a high-dimensional projection space and multi-crop augmentation. The framework produced strong representations on ImageNet and exhibited emergent properties such as object segmentation through attention maps. Subsequent work refined the framework: DINOv2~\cite{oquab2024dinov2} scaled the approach to curated image datasets at billion-parameter scale, and register tokens~\cite{darcet2024registers} addressed feature-map artifacts that emerge in self-supervised ViTs. We adopt the canonical DINO formulation with register tokens as a stability mechanism for the Vision Transformer backbones.

\paragraph{Self-distillation in audio.}
Two prior works have applied self-distillation in audio sub-domains. DinoSR~\cite{liu2023dinosr} combines self-distillation with masked language modeling and online clustering, targeting speech representation learning on LibriSpeech. ASiT~\cite{atito2024asit} extends DINO's self-distillation with group-masked model learning for audio event classification, hybridizing self-distillation with masked spectrogram modeling. Both modify the canonical DINO framework: DinoSR adds clustering and masking targeted at speech, ASiT adds group-masked modeling to capture local statistical structure of spectrograms. Neither performs a direct adaptation of canonical DINO to general audio under the BYOL-A pattern, nor a direct comparison against the latent-prediction paradigm under matched conditions. DINO-A occupies this gap.

\section{Method}
\label{sec:method}

DINO-A adapts the self-distillation framework of DINO~\cite{caron2021dino} to general audio representation learning, following the adaptation pattern established by BYOL-A~\cite{niizumi2021byola,niizumi2023byola2}, which transferred BYOL~\cite{grill2020byol} from vision to audio. We retain the canonical components of DINO and replace only the input modality and augmentation pipeline. This section summarizes the framework and details the audio-specific design choices.

\subsection{Self-Distillation Framework}
\label{sec:method:dino}

DINO trains a student network $g_{\theta_s}$ to match the output distribution of a teacher network $g_{\theta_t}$, which shares the same architecture and whose parameters are an exponentially-moving-average (EMA) of the student parameters. Given an input $x$, two global views and several local views are generated through augmentation. The student processes all views, while the teacher processes only the global ones. Both networks output probability distributions over a high-dimensional projection space (here, $K = 65{,}536$ dimensions). The loss is the cross-entropy between teacher and student distributions, summed over all student-teacher view pairs:

\begin{equation}
\mathcal{L}_{\mathrm{DINO}} = - \sum_{x_g} \sum_{\substack{x' \in V \\ x' \neq x_g}} P_t(x_g)^\top \log P_s(x'),
\end{equation}

where $V$ is the set of all views, $x_g$ ranges over global views, and $P_t, P_s$ denote softmax outputs of teacher and student, respectively. To prevent collapse, the teacher output is centered (subtracting an EMA of past outputs) and sharpened with a low teacher temperature, while the student uses a higher temperature; specific values are given in Section~\ref{sec:setup:pretrain-cfg}.

\subsection{Audio-Specific Augmentations}
\label{sec:method:aug}

Standard DINO augmentations (color jittering, blurring, solarization) are designed for natural images and do not transfer meaningfully to spectrograms. We instead adopt the augmentation pipeline of BYOL-A v2~\cite{niizumi2023byola2}, which was specifically designed for log-mel spectrograms. Each view is generated by sequentially applying:

\begin{itemize}
    \item \textbf{Mixup-BYOLA}: a log-exp-domain mixup that blends the current spectrogram with a randomly sampled one from a memory bank of size 2048, with mixing ratio $\alpha = 0.4$. The blending is performed in the linear-amplitude domain after exponentiating the log-mel input, then converted back to log scale.
    \item \textbf{RandomResizeCrop}: a random crop in the time-frequency plane followed by resizing back to $64 \times 96$, applied independently to each view.
\end{itemize}

After augmentation, post-normalization is applied across the batch using running statistics, following the RunningNorm scheme of BYOL-A v2.

\subsection{Multi-Crop Strategy}
\label{sec:method:multicrop}

We follow the standard DINO multi-crop setup with two global views (sized $1.0 \times$ to $1.5 \times$ the canonical input) and six local views (sized $0.5 \times$ to $1.0 \times$). Both view types operate on the same 0.95-second audio segment. The student processes all eight views, while the teacher processes only the two global views.

The semantic appropriateness of multi-crop in audio is less obvious than in vision: a small temporal-frequency crop may capture a transient or silence rather than a coherent part of the underlying sound event. We discuss the implications of this design choice in Section~\ref{sec:discussion:multicrop}.

\subsection{Backbone Architectures}
\label{sec:method:backbone}

We evaluate three backbone architectures, all sharing the same DINO loss, augmentation pipeline, and EMA teacher schedule:

\paragraph{AudioViT (8$\times$8 patches).} A Vision Transformer~\cite{dosovitskiy2021vit} with patch size $8 \times 8$ on the $64 \times 96$ input, yielding $8 \times 12 = 96$ patch tokens. Embedding dimension is 384, with a standard transformer encoder. We additionally include four register tokens~\cite{darcet2024registers}, which have been shown to stabilize attention maps in self-supervised ViTs.

\paragraph{AudioViT (16$\times$16 patches).} Identical to the above but with patch size $16 \times 16$, producing $4 \times 6 = 24$ patch tokens. This configuration tests whether the lower spatial resolution affects the quality of learned audio representations.

\paragraph{AudioNTT2022 (CNN).} The convolutional encoder used as the BYOL-A v2 backbone~\cite{niizumi2023byola2}. We include it as a non-Transformer baseline within the same DINO framework, isolating the effect of the architectural family from the self-supervised objective.

\section{Experimental Setup}
\label{sec:setup}

\subsection{Pretraining Dataset}
\label{sec:setup:pretrain}

We pretrain DINO-A on FSD50K~\cite{fonseca2022fsd50k}, an open dataset of 51{,}197 audio clips collected from Freesound and labeled with 200 classes drawn from the AudioSet ontology~\cite{gemmeke2017audioset}. Clips range from 0.3 to 30 seconds in length and cover a broad acoustic spectrum including speech, music, environmental sounds, animals, and mechanical noises. We use the development split (approximately 41{,}000 clips, 80 hours of audio) for self-supervised pretraining; no labels are used at any stage.

The choice of FSD50K is motivated by two considerations. First, it enables direct comparison with BYOL-A v2~\cite{niizumi2023byola2}, which uses FSD50K as its default pretraining corpus. Pretraining DINO-A on the same dataset isolates the effect of the self-supervised algorithm from the effect of pretraining data. Second, FSD50K is computationally feasible on a single consumer-grade GPU, while AudioSet-scale pretraining would require multi-GPU clusters or training times measured in weeks.

\subsection{Audio Preprocessing}
\label{sec:setup:preproc}

All audio is resampled to 16~kHz mono. From each clip, a 0.95-second segment (15{,}200 samples) is randomly cropped at every training iteration. Spectrograms are computed with FFT size 1024, hop length 160, and 64 mel bins covering 60--7800~Hz, yielding a $64 \times 96$ log-mel representation. Augmentations are applied as described in Section~\ref{sec:method:aug}.

\subsection{Pretraining Configuration}
\label{sec:setup:pretrain-cfg}

We train each backbone for 100 epochs with batch size 100 using AdamW. The peak learning rate is $5 \times 10^{-4}$ scaled linearly to the training batch size, $\mathrm{lr}_{\mathrm{peak}} = 5 \times 10^{-4} \cdot (100/256) \approx 1.95 \times 10^{-4}$. The schedule consists of linear warmup from $0$ to $\mathrm{lr}_{\mathrm{peak}}$ over the first 10 epochs, followed by cosine decay from $\mathrm{lr}_{\mathrm{peak}}$ to $10^{-6}$ over the remaining 90 epochs. Weight decay follows a cosine schedule from 0.04 to 0.4. Teacher temperature is held constant at $\tau_t = 0.04$, student temperature at $\tau_s = 0.1$. The EMA teacher momentum follows a cosine schedule from 0.996 to 1.0. Gradients are clipped to a maximum norm of 3.0. The DINOHead projection has output dimension $K = 65{,}536$ and uses batch normalization with normalized last layer. Multi-crop generates 2 global views and 6 local views per training sample. Pretraining runs on a single NVIDIA RTX 5090 GPU and completes within 3 to 5 days depending on backbone size.

\subsection{Downstream Evaluation}
\label{sec:setup:eval}

We evaluate the pretrained encoders using the linear probing protocol of EVAR~\cite{niizumi2022composing}\footnote{Implementation: \url{https://github.com/nttcslab/eval-audio-repr}.}, which freezes the encoder and trains a linear classifier on top of the extracted representations. This evaluation isolates the quality of self-supervised features without confounding effects from end-to-end fine-tuning.

We report results on four standard audio classification benchmarks covering distinct acoustic domains:

\begin{itemize}
    \item \textbf{ESC-50}~\cite{piczak2015esc50}: 2{,}000 environmental sound clips across 50 classes.
    \item \textbf{Speech Commands v2 (SPC-v2)}~\cite{warden2018speechcommands}: 105{,}829 one-second utterances of 35 spoken commands.
    \item \textbf{UrbanSound8K (US8K)}~\cite{salamon2014urbansound}: 8{,}732 urban sound clips across 10 classes.
    \item \textbf{GTZAN}~\cite{tzanetakis2002gtzan}: 1{,}000 music clips across 10 genres.
\end{itemize}

The same evaluation pipeline is applied identically to all DINO-A variants and to the BYOL-A v2 baseline, ensuring direct comparability of results.

\section{Results}
\label{sec:results}

\subsection{Main Results}
\label{sec:results:main}

Table~\ref{tab:main} reports linear probing accuracy for the three DINO-A variants and the BYOL-A v2 baseline. The following subsections analyze patch resolution effects, CNN-vs-ViT trade-offs, and the gap to BYOL-A v2.

\begin{table}[H]
\centering

\caption{Linear probing accuracy (\%) on four downstream audio classification tasks. All models are pretrained on FSD50K with no labels and evaluated with frozen backbone using the EVAR protocol. Best DINO-A variant per column underlined.}
\label{tab:main}
\begin{tabular}{lccccc}
\toprule
Model & ESC-50 & SPC-v2 & US8K & GTZAN & Average \\
\midrule
BYOL-A v2~\cite{niizumi2023byola2} & 83.20 & 91.81 & 79.65 & 73.44 & \textbf{82.03} \\
\midrule
DINO-A ViT (8$\times$8) & \underline{72.50} & 85.93 & \underline{73.20} & \underline{48.66} & \underline{70.07} \\
DINO-A ViT (16$\times$16) & 71.60 & 85.62 & 72.25 & 48.00 & 69.37 \\
DINO-A CNN (AudioNTT2022) & 60.40 & \underline{88.56} & 65.86 & 45.32 & 65.03 \\
\bottomrule
\end{tabular}
\end{table}

\subsection{Patch Resolution in Audio ViTs}
\label{sec:results:patch}

The two AudioViT variants differ only in patch size (8$\times$8 vs.\ 16$\times$16) while sharing all other hyperparameters. The 8$\times$8 configuration outperforms 16$\times$16 by 0.7pp on average, with consistent advantages on ESC-50 (+0.9pp), SPC-v2 (+0.3pp), and US8K (+0.95pp). The smaller patches yield 96 patch tokens compared to 24 for the larger ones, providing finer-grained spatial decomposition of the spectrogram.

This pattern aligns with observations from the original DINO paper~\cite{caron2021dino}, which reports that smaller patches improve representation quality in vision. Our results are consistent with the same trend holding for log-mel spectrograms, suggesting that fine-grained tokenization of the time-frequency plane may be beneficial for self-distillation in the audio domain.

\subsection{CNN versus ViT Trade-offs}
\label{sec:results:cnn-vit}

The CNN backbone trails ViT 8$\times$8 by 5.04pp on average, but the per-task pattern is informative. On Speech Commands v2, the CNN actually outperforms both ViT variants by approximately 2.6--2.9pp, reaching 88.56\% accuracy. On the other three tasks, the CNN underperforms substantially, falling 12.1pp behind on ESC-50 and 7.3pp on US8K.

The pattern is consistent with the inductive biases of the two architectures. CNNs encode translation equivariance and local connectivity, which match the structure of speech where short-time spectral patterns (formants, phonemes) carry most of the discriminative information. Vision Transformers lack such inductive biases but compensate through global self-attention, which appears to be more useful for environmental sounds and music where coarse spectral context matters more than local detail.

\subsection{Comparison with BYOL-A v2 Baseline}
\label{sec:results:vs-baseline}

Across all four downstream tasks, BYOL-A v2 outperforms the best DINO-A variant. The average gap is 11.96pp (82.03\% vs.\ 70.07\%), with the largest deficit on GTZAN (24.78pp) and the smallest on SPC-v2 (5.88pp). Notably, the gap is not uniform: on speech-heavy SPC-v2, the CNN-based DINO-A reaches 88.56\%, narrowing the gap to 3.25pp.

The most plausible interpretation is that DINO is less data-efficient than BYOL in the audio domain at the FSD50K scale: the cross-entropy loss over 65{,}536 output dimensions in the DINOHead provides a more granular learning signal than BYOL-A's cosine similarity in a lower-dimensional projection, but requires larger and more diverse pretraining data to populate that signal meaningfully. A secondary contributing factor is that the multi-crop self-distillation objective relies on assumptions about local-global consistency that hold less robustly for audio than for natural images. We elaborate on both factors in Section~\ref{sec:discussion}.

\section{Discussion}
\label{sec:discussion}

The empirical results of Section~\ref{sec:results} establish that DINO-A produces useful audio representations across diverse downstream tasks but trails BYOL-A v2 on average. This section examines the most plausible factors driving the gap and the architectural patterns observed within the DINO-A family. We organize the discussion around three questions: why DINO-A underperforms BYOL-A v2 on FSD50K, why ViT and CNN backbones behave differently across task types, and what these findings imply for future audio self-supervised learning research.

\subsection{The Pretraining Scale Hypothesis}
\label{sec:discussion:scale}

The most likely explanation for DINO-A's gap relative to BYOL-A v2 lies in the interaction between the DINO loss and the size of the pretraining corpus. The DINOHead projects representations into 65{,}536 dimensions and trains the student to match a teacher distribution over this space via cross-entropy. This design provides a rich learning signal: each output dimension acts as a soft pseudo-class that captures some aspect of the input. With FSD50K's 41{,}000 pretraining clips, the ratio of training examples to output dimensions is below 1, which means each output dimension receives statistically sparse supervision per epoch. Centering and sharpening compensate partially, but the signal remains less informative than at the data-to-output ratios at which DINO was originally validated. ImageNet's training-examples-to-output-dimensions ratio (1.3M / 65{,}536 $\approx$ 20) is approximately 30 times that of FSD50K (41k / 65{,}536 $\approx$ 0.6).

BYOL-A v2 operates under a fundamentally different regime. Its loss is the cosine similarity between projected representations of two augmented views, computed in a low-dimensional projection space. This objective makes weaker statistical demands on the data: it requires only that the encoder map augmented views of the same input to nearby points, without the auxiliary structure of a high-dimensional classification target. As a result, BYOL-A v2 is expected to be more data-efficient at small to medium pretraining scales, while DINO is expected to benefit more from scaling.

This hypothesis is consistent with the broader audio SSL literature. Methods that depend on rich, fine-grained learning signals, such as masked spectrogram reconstruction in Audio-MAE or discrete acoustic token prediction in BEATs, are typically pretrained on AudioSet (approximately 2 million clips) rather than FSD50K. Methods based on latent prediction with smaller projection spaces, such as BYOL-A, achieve competitive performance on FSD50K-scale corpora. Our results are consistent with DINO falling on the data-hungry side of this dichotomy, though direct scaling experiments are required for confirmation.

\subsection{Multi-Crop Semantics in Audio}
\label{sec:discussion:multicrop}

A second contributing factor concerns the structural assumptions of multi-crop augmentation. The local-to-global prediction task that DINO uses requires that local views carry information consistent with the global representation, so that the student can map a partial observation to the full one. In images, this assumption is supported not only by the presence of object parts in local crops but also by the textural and chromatic regularities of natural scenes: even an uninformative crop of background sky or grass remains globally consistent with the rest of the image at the level of low-level statistics.

Audio spectrograms lack this redundant low-level structure. A short time-frequency crop of a log-mel spectrogram either captures the relevant acoustic event or does not, with limited intermediate cases. Sounds with low temporal density, transient onsets, or rapid spectral variation produce local views whose statistics differ qualitatively from the global view rather than gradually. The student-teacher consistency objective is then asked to enforce alignment across views that may be only weakly related, which is a noisier learning signal than its vision counterpart.

This concern is specific to DINO-A, since BYOL-A v2 does not use multi-crop and instead trains with two augmented views of equivalent size. The local-to-global prediction task is therefore an additional structural cost that DINO-A bears but BYOL-A v2 does not, compounding rather than paralleling the scale issue. Larger and more diverse pretraining corpora mitigate the effect statistically, since the proportion of degenerate local views shrinks as the variety of represented acoustic events grows. This complements the scale hypothesis of Section~\ref{sec:discussion:scale}: both arguments converge on AudioSet-scale pretraining as a natural direction for closing the gap to BYOL-A v2.

\subsection{Architectural Trade-offs}
\label{sec:discussion:arch}

The patch resolution comparison in Section~\ref{sec:results:patch} confirms a known property of vision transformers in the audio context: smaller patches yield finer-grained tokenization and consistently better representations. The 8$\times$8 configuration produces 96 tokens with sufficient temporal-frequency resolution to capture short-time spectral patterns relevant to most downstream tasks, while 16$\times$16 with 24 tokens averages over time-frequency regions large enough to obscure such patterns.

The CNN versus ViT comparison reveals a sharper trade-off. On Speech Commands v2, the CNN backbone outperforms both ViT variants, while on the other three tasks it underperforms substantially. This pattern reflects the inductive biases of the two families. Speech Commands consists of one-second utterances of short spoken keywords, where discriminative information lies in localized spectral patterns: the formant transitions and consonant-vowel boundaries that occupy specific time-frequency regions. A CNN's local receptive fields and translation equivariance match this structure directly. Vision Transformers, lacking such inductive biases, must learn equivalent regularities from data, which is harder under the limited pretraining scale of FSD50K. On environmental sounds (ESC-50, US8K) and music (GTZAN), the relevant features are more globally distributed across the spectrogram, and the global self-attention of ViT becomes advantageous despite its weaker priors.

This finding suggests that backbone choice for audio self-distillation should be task-dependent: CNN for speech-heavy applications, ViT with small patches for environmental and musical audio. A hybrid backbone that combines local inductive biases with global attention may offer a compelling middle ground, though we leave such a study to future work.

\subsection{Limitations and Future Work}
\label{sec:discussion:limitations}

Three limitations of the present study merit explicit acknowledgment. First, our pretraining is restricted to FSD50K. Confirming the scale hypothesis requires pretraining DINO-A on AudioSet and observing whether the gap to BYOL-A v2 closes; such experiments are computationally expensive and remain ongoing. Second, our evaluation uses only linear probing on frozen features. Full fine-tuning may reveal a different ranking, particularly for backbones whose features are linearly suboptimal but contain useful nonlinear structure. Third, we adopt the BYOL-A augmentation pipeline without modification. Audio-native augmentations such as SpecAugment, pitch shifting, or time stretching may interact differently with the multi-crop self-distillation framework, and a systematic study of audio-specific augmentations for DINO-A would be a natural extension.

Beyond the canonical 8$\times$8 and 16$\times$16 square patches tested here, the asymmetric structure of log-mel spectrograms motivates exploring non-square patch geometries. The time and frequency axes of a spectrogram have qualitatively different units (linear time vs.\ logarithmic mel-frequency) and carry different types of acoustic information, so a square patch is a default inherited from natural-image transformers rather than a design optimal for audio. Prior work on supervised audio classification reports that strongly anisotropic patches such as full-frequency-by-narrow-time configurations outperform square ones when training from scratch~\cite{gong2021ast}, indicating that the asymmetry can be exploited at the input representation level. How such patch geometries interact with DINO's multi-crop self-distillation is not obvious, since extreme anisotropy may conflict with the multi-crop assumption that local views retain meaningful 2D structure. Mild non-square patches, such as 4$\times$8 or 8$\times$4, may offer a sweet spot worth testing. We identify this as a natural extension of the present work.

\section{Conclusion}
\label{sec:conclusion}

This work presented DINO-A, an adaptation of the DINO self-distillation framework to general audio representation learning, following the BYOL-to-BYOL-A adaptation pattern. Under identical pretraining on FSD50K and identical linear probing evaluation, DINO-A produces useful audio representations across four downstream benchmarks but trails BYOL-A v2 by 11.96 percentage points on average. Architectural ablations within the DINO-A family reveal that smaller patches (8$\times$8) consistently outperform larger ones (16$\times$16), and that the choice between Vision Transformer and convolutional backbone interacts strongly with the type of downstream task: CNNs win on speech, ViTs win on environmental sounds and music. We trace the gap to BYOL-A v2 primarily to the interaction between DINO's high-dimensional projection space and FSD50K's limited scale, with multi-crop augmentation introducing an additional structural cost that DINO-A bears but BYOL-A v2 does not. Both factors point toward AudioSet-scale pretraining as the natural next step for closing the gap. Beyond scale, we identify hybrid objectives, audio-native augmentations, and asymmetric patch geometries as promising directions for further investigation.

\section*{Acknowledgments}
The authors used a large language model (Claude, Anthropic) for manuscript drafting, language editing, and citation retrieval. All scientific content is the authors' own work.

\bibliographystyle{plain}
\bibliography{references}

\end{document}